\documentclass[twocolumn,aps,prb,showpacs,amsmath,amssymb,superscriptaddress,bibnotes,longbibliography]{revtex4-2} 
\usepackage{hyperref}
\usepackage[usenames,dvipsnames]{color}
\usepackage{amsmath}
\usepackage{amssymb}
\usepackage{amsthm}
\usepackage{graphicx}
\usepackage{epstopdf}
\usepackage{todonotes}
\usepackage[normalem]{ulem}
\usepackage{verbatim}
\usepackage{stmaryrd}
\usepackage{lipsum}
\usepackage{mathtools}
\usepackage[export]{adjustbox}

\newcommand{\m}{\text{-}}

\newcommand{\ua}{\uparrow}
\newcommand{\da}{\downarrow}

\begin{document}

\title{Mechanism of Superconductivity in the Hubbard Model at Intermediate Interaction Strength}
\author{Xinyang Dong}
\affiliation{Department of Physics, University of Michigan, Ann Arbor, MI 48109, USA}
\author{Lorenzo Del Re}
\affiliation{Department of Physics, Georgetown University, NW, Washington, DC 20057, USA}
\affiliation{Max Planck Institute for Solid State Research, D-70569 Stuttgart, Germany}
\author{Alessandro Toschi}
\affiliation{Institute of Solid State Physics, TU Wien, 1040 Vienna, Austria}
\author{Emanuel Gull}
\affiliation{Department of Physics, University of Michigan, Ann Arbor, MI 48109, USA}

\begin{abstract}
    We study the fluctuations responsible for pairing in the $d$-wave superconducting state of the two-dimensional Hubbard model at intermediate coupling within a cluster dynamical mean-field theory with a numerically exact quantum impurity solver. By analyzing how momentum and frequency dependent fluctuations generate the $d-$wave superconducting state in different representations, we identify antiferromagnetic  fluctuations as the pairing glue of superconductivity both in the underdoped and the overdoped regime. 
    Nevertheless, in the intermediate coupling regime, the predominant magnetic fluctuations may differ significantly from those described by conventional spin-fluctuation theory.
\end{abstract}

\maketitle
\makeatletter
\let\toc@pre\relax
\let\toc@post\relax
\makeatother

\section{Introduction}
The microscopic mechanism of unconventional high-temperature superconductivity has been one of the most controversially debated topics in condensed matter physics since the discovery of superconductivity in layered copper-oxides in 1986. While several aspects of the observed physics, such as the $d-$wave symmetry of the order parameter and the proximity to an antiferromagnetic Mott phase, clearly suggest that  superconductivity must emerge from strongly correlated electronic processes, the intrinsic quantum many-body nature of the problem has hitherto prevented a rigorous identification of the pairing glue.
To explicitly address this point, we present a focused study of the origin of superconductivity in the two dimensional single band Hubbard model.
The Hubbard Hamiltonian, which includes a kinetic term describing the hopping between neighbouring sites on a lattice and a potential-energy term encoding a local electrostatic repulsion, is a minimal theoretical model believed to capture the salient aspects of cuprate superconductivity.

Among the theoretical explanations proposed for the origin of the high-temperature superconductivity in this context, spin fluctuations have been a prominent scenario since the beginning \cite{Scalapino1986, Miyake1986, Scalapino1999, Scalapino2012}.
In particular, within in the weak-coupling regime of the Hubbard model, renormalization group techniques \cite{ Halboth2000_PRL, Zanchi2000,Honerkamp2001, Raghu2010} find $d$-wave superconductivity in qualitative agreement with spin-fluctuation exchange studies \cite{Raghu2010, Scalapino1986}, consistent with diagrammatic Monte Carlo calculations \cite{Deng2015}. 
At the same time,  other, qualitatively different microscopic pictures of superconductivity exist besides the spin fluctuations, including the RVB theory \cite{Anderson1987}, nematic fluctuations \cite{Fradkin2010}, loop current order \cite{Allais2012}, or the ``intertwining'' of  orders of different types \cite{Fradkin2015}.
In fact, to what extent the weak-coupling spin fluctuation results apply to the much stronger interaction values, which are typical of cuprate materials, and whether there are other competing or intertwining fluctuations driving the superconductivity remains unresolved.

To provide a conclusive answer, we perform an analysis of the anomalous self energy in the $d-$wave superconducting state within the  method of fluctuation diagnostics \cite{Gunnarsson2015}.
We note that, unlike other diagrammatic approaches, which postulate a specific physical mechanism, analyze its consequences, and then compare to experiments, the fluctuation diagnostics procedure treats fluctuations of all kinds, including those possibly driving superconductivity, on  equal footing,  and is applicable in all parameter regimes, independent of the degree of correlation.
However, the fluctuation diagnostics procedure as derived in \cite{Gunnarsson2015} was  only applicable to the highly symmetric normal state, and thus cannot be used to analyze superconductivity. Hence, we will first generalize this approach to  the case of phases with spontaneously broken symmetries and then apply it to identify the dominant fluctuations driving the anomalous self energy in the superconducting state.

\section{Method}

The Hamiltonian of the two-dimensional single band Hubbard model is
\begin{align}
    H=\sum_{k\sigma}(\varepsilon_k-\mu)c^\dagger_{k\sigma}c_{k\sigma}+U\sum_in_{i\uparrow}n_{i\downarrow}, \label{eq:Hubbard}
\end{align}
with $i$ a lattice site, $k$ momentum, $c^{(\dagger)}$ annihilation (creation) operators, and $n$ the density. $\varepsilon_k=-2t(\cos k_x + \cos k_y)$ is the dispersion with hopping $t$, $U$ the interaction strength, and $\mu$ the chemical potential.
We use the dynamical cluster approximation (DCA) \cite{Maier2005} on a cluster with size $N_c=8$ with a numerically exact continuous time auxiliary field (CTAUX) \cite{Gull2008,Gull2011} impurity solver to enter the superconducting state non-perturbatively \cite{Gull2013}, and obtain Green's functions, self-energies, and vertex functions. 

In the DCA, the momentum structure of the Hubbard model self-energy is approximated by $N_c$ basis functions which retain the full frequency dependence \cite{Maier2005}. To enter the superconducting phase, we allow for order and provide a superconducting bias field at the first iteration, removing it in subsequent iterations to converge to the equilibrium state. Within the $8$-site DCA approximation, the model exhibits a large and stable $d$-wave superconducting region \cite{Chen2015,Gull2013}. The model is known to also exhibit a stripe phase \cite{Zheng2017, Huang2018, Qin2020, Mai2021, Wietek2021}, to which our calculation is not sensitive, since its periodicity is larger than the $N_c=8$ cluster.

To identify the superconducting glue, we apply the fluctuation diagnostics scheme \cite{Gunnarsson2015} to the anomalous self-energy in the superconducting state. This approach, which so far has  been derived \cite{Gunnarsson2015} and applied \cite{Gunnarsson2015,Schaefer2021}  only in the paramagnetic normal state, allows for a rigorous identification of the dominant scattering mechanisms responsible for the observed self-energy.
Fluctuation diagnostics exploits symmetries in the Hamiltonian that lead to different expressions for the Schwinger-Dyson (SD) equation of the self energy $\Sigma$
\begin{align}
   \label{eq:SD} 
   \Sigma(18) - \Sigma_{\infty} = - U(1234) G(25) F(5678) G(63) G(74) \, , 
\end{align}
where we use the Einstein summation notation and introduce the short-hand notation $i = (K_i, \sigma_i, \tau_i)$ for momentum, spin and time indices. The two indices on the self energy $\Sigma(12)$ and the Green's functions $G(12)$ represent their normal ($1 = 2$) and anomalous ($1 = -2$) components, with $-i = (-K_i, -\sigma_i, \tau_i)$ . $\Sigma_{\infty}$ is the static Hartree contribution for the normal self energy, and $F$ is the full two-electron scattering amplitude.
$U(1234)$ denotes the antisymmetrized interaction which, in the Hubbard model, is proportional to the local interaction $U$.
This expression is exact and relates two-particle fluctuations to single-particle quantities.

\begin{figure*}[tb]
    \centering
    \includegraphics[width=0.9\textwidth]{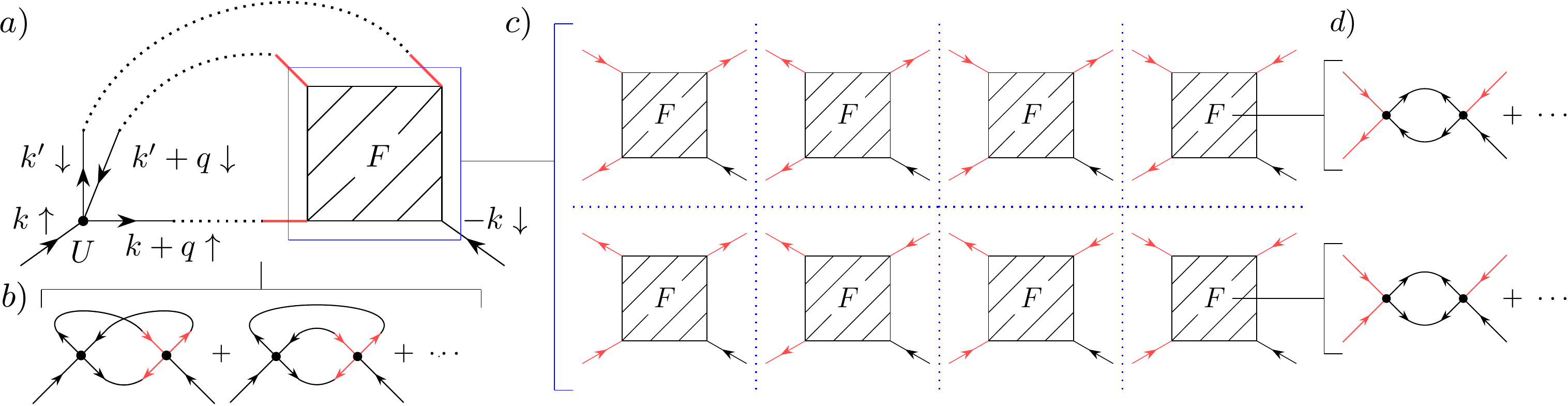}
    \caption{Anomalous self-energy diagrams. Panel a: Schwinger Dyson equation (Eq.~(\ref{eq:SD})). Shaded box denotes vertex $F$ with one fixed outgoing leg. Dotted lines: normal or anomalous Green's function. Panel b: Low-order anomalous self energy diagrams. Panel c: Explicit representation of index combinations of $F$. Panel d: Some of the low-order diagrams contributing to $F$.}
    \label{fig:diagram}
\end{figure*}
%
%
Panel (a) of Fig.~\ref{fig:diagram} shows a diagrammatic representation of the anomalous self-energy of Eq.~\ref{eq:SD} in frequency space, with $k=(K, i\omega_n)$ representing fermionic and $q = (Q, i\nu_n)$ representing bosonic indices. A choice of $k_1 = k$, $\sigma_1 = \uparrow$, $k_2 = k+q$, $ \sigma_2 = \uparrow$, $k_3 = k'+q$, $\sigma_3 = \downarrow$, $k_4 = k'$, $\sigma_4 = \downarrow$, and $k_8 = -k$, $\sigma_8 = \downarrow$ satisfies momentum, energy, and spin conservations.
Panel (b) shows two low-order terms.
$F$, in the case of superconducting order, can have eight possible combinations of incoming or outgoing legs, as illustrated in panel (c). It then contains all allowed scattering processes, some of which are illustrated in panel (d).

The different expressions of the SD are equivalent when all internal $k$ summations are performed.
Important information about the role played by the different scattering channels can be gained by comparing the expressions of the SD equation after \textit{partial} summations over the internal variables $k$.

From a physical point of view, each expression can be associated to one of the possible collective modes (e.g., density, magnetic, singlet/triplet pairing) of the electronic system.
A large contribution to the final sum over $q$ at low transfer frequency and at a definite momentum signifies a dominant collective mode, in contrast to contributions more evenly distributed  over a wide range of frequencies and momenta.

We now derive Eq.~\ref{eq:SD} in more detail. To have access to all the single- and two- particle quantities in the SD equation, we introduce the Bogoliubov-de Gennes spinors in momentum space \cite{Zhu2016}
%
\begin{align}
    \Phi_{K}=\begin{pmatrix}
    c_{K\uparrow}&c_{K\downarrow}&
    c^\dagger_{-K\uparrow}&c^\dagger_{-K\downarrow}
    \end{pmatrix}^T \, .
    %
\end{align}
\label{eq:bdg_spinor}
%
The single-particle Green's function in the singlet superconducting state is
\begin{align}
\label{eq:GF_BdG}
    &\underline{\mathcal{G}}_{K}(\tau) 
    = -\langle \mathcal{T} \Phi_{K}(\tau) \Phi_{K}^{\dagger}(0) \rangle \\
    &= -\left\langle \mathcal{T} 
    \begin{pmatrix} 
    c_{K\uparrow}c^\dagger_{K\uparrow}
    &\hspace{-0.15cm}0
    &\hspace{-0.15cm}0
    &\hspace{-0.15cm}c_{K\uparrow}c_{\text{-}K\downarrow} 
    \\
    0
    &\hspace{-0.15cm}c_{K\downarrow}c^\dagger_{K\downarrow}
    &\hspace{-0.15cm}c_{K\downarrow}c_{\text{-}K\uparrow}
    &\hspace{-0.15cm}0
    \\
    0
    &\hspace{-0.15cm}c^\dagger_{\text{-}K\uparrow}c^\dagger_{K\downarrow}
    &\hspace{-0.15cm}c^\dagger_{\text{-}K\uparrow}c_{\text{-}K\uparrow}
    &\hspace{-0.15cm}0
    \\
    c^\dagger_{\text{-}K\downarrow}c^\dagger_{K\uparrow}
    &\hspace{-0.15cm}0 
    &\hspace{-0.15cm}0
    &\hspace{-0.15cm}c^\dagger_{\text{-}K\downarrow}c_{\text{-}K\downarrow}
    \end{pmatrix}(\tau, 0)
    \right\rangle \, , \nonumber
\end{align}
where ${\cal T}$ is the (imaginary) time-ordering operator.
SU(2) symmetry reduces the number of independent terms in Eq.~\ref{eq:GF_BdG} to four, such that Eq.~\ref{eq:GF_BdG} can be written in a compact form as
\begin{align}
\underline{\mathcal{G}}_K(\tau) &= 
- \langle \mathcal{T} \begin{pmatrix} 
c_{K\uparrow}(\tau)c^\dagger_{K\uparrow}(0) &  c_{K\uparrow}(\tau)c_{-K\downarrow}(0) \\
c^\dagger_{-K\downarrow}(\tau)c^\dagger_{K\uparrow}(0) &  c^\dagger_{-K\downarrow}(\tau)c_{-K\downarrow}(0)
\end{pmatrix} \rangle \nonumber \\[0.3em]
&= - \begin{pmatrix} 
G_K^N(\tau) &  G_K^A(\tau) \\[0.3em]
G_K^{A \dagger}(\tau) &  -G_{-K}^N(-\tau)
\end{pmatrix} \, .
\end{align}
The single particle Green's function matrix contains both normal (N) and anomalous (A) entries, with the normal terms defined as $G^{\sigma\sigma}_{K}(\tau) = -\langle \mathcal{T} c_{K\sigma}(\tau) c_{K\sigma}^\dagger(0) \rangle$, $\sigma = \uparrow, \downarrow$, and the anomalous terms defined as $G^A_{K}(\tau) = G^{\uparrow\downarrow}_{K}(\tau) = -\langle \mathcal{T} c_{K \uparrow}(\tau) c_{-K\downarrow}(0) \rangle$,  $G^{A\dagger}_{K}(\tau) = G^{\downarrow\uparrow}_{K}(\tau) = -\langle \mathcal{T} c_{-K \downarrow}^\dagger(\tau) c_{K\uparrow}^\dagger(0) \rangle$.
Fourier transforming to frequency space and introducing short hand notations $k=(K, i\omega_n)$,
we can define $G^N_{k} = G^{\ua\ua}_{k} = G^{\da\da}_{k}$. 
For d-wave superconductivity on a lattice with inversion symmetry, the anomalous Green's function can be chosen to be real \cite{Lichtenstein2000}, such that $G^A_{\pm k} = G^{A\dagger}_{\pm k}$.
The self energy can then be computed with
\begin{align}
    \underline{\Sigma}(k) =  \underline{\mathcal{G}}^{-1}_{0}(k) - \underline{\mathcal{G}}^{-1}(k) \, ,
\end{align}
where
\begin{align}
    \underline{\Sigma}(k) &= 
    \begin{pmatrix} 
    \Sigma^N_{k\uparrow} &\Sigma^A_{k\uparrow} \\[0.3em]
    \Sigma^{A*}_{k\uparrow} &-\Sigma^N_{-k\downarrow}
    \end{pmatrix} \, ,
    \\
    \underline{\mathcal{G}}_0^{-1}(k) &=
    \begin{pmatrix} 
    i\omega_n - \varepsilon_k + \mu &0 \\
    0 & i\omega_n + \varepsilon_k - \mu
    \end{pmatrix} \, .
\end{align}

The two-particle Green's function takes the form
\begin{align}
    G^{(2)}(1 2 3 4) = \langle \mathcal{T} o_1 o_2 o_3 o_4\rangle \, ,
\end{align}
where we use $i$ as a short-hand notation for momentum, spin and imaginary time indices $(K_i, \sigma_i, \tau_i)$. $o_i$ is either a creation operator $c_i^\dagger$ or an annihilation operator $c_i$. 

In the paramagnetic state, the number of creation and annihilation operators in the two-particle Green's function must be equal to preserve charge conservation.
In the superconducting state, the broken U(1) symmetry gives in total $2^4=16$ combinations of different creation or annihilation operators, which can be written in matrix form as
\begin{align}
    &\underline{\mathcal{G}}^{(2)}(1234) \\
    &=
    \left\langle
    \mathcal{T}
    \begin{pmatrix}
    c_1^\dagger c_2 c_3^\dagger c_4 
    & c_1^\dagger c_{2}c_{\m3}c_4
    &c_{1}^\dagger c_2 c_3^\dagger c_{\m4}^\dagger 
    & c_{1}^\dagger c_{2} c_{\m3} c_{\m4}^\dagger \\
    c_1^\dagger c_{\m2}^\dagger c_{3}^\dagger c_4
    & c_1^\dagger c_{\m2}^\dagger c_{\m3} c_4 
    &c_{1}^\dagger c_{\m2}^\dagger c_{3}^\dagger c_{\m4}^\dagger 
    & c_{1}^\dagger c_{\m2}^\dagger c_{\m3} c_{\m4}^\dagger \\
    c_{\m1}c_2 c_3^\dagger c_4 
    & c_{\m1}c_{2}c_{\m3}c_{4}
    & c_{\m1} c_2 c_3^\dagger c_{\m4}^\dagger 
    &c_{\m1} c_{2} c_{\m3} c_{\m4}^\dagger \\
    c_{\m1} c_{\m2}^\dagger c_{3}^\dagger c_{4} 
    & c_{\m1} c_{\m2}^\dagger c_{\m3} c_{4}
    & c_{\m1} c_{\m2}^\dagger c_{3}^\dagger c_{\m4}^\dagger 
    & c_{\m1} c_{\m2}^\dagger c_{\m3} c_{\m4}^\dagger
    \end{pmatrix}
    \right\rangle \, , \nonumber
\end{align}
with $i = (K_i, \sigma_i, \tau_i)$, $-i = (-K_i, -\sigma_i, \tau_i)$. 
\\
Each term $G^{(2)}(1234)$ in the two-particle Green's function matrix can be decomposed into connected ($G_c^{(2)}$) and disconnected parts as
\begin{align}
    G^{(2)}_c(1234) &=  G^{(2)}(1 2 3 4) - \langle \mathcal{T} o_1 o_2\rangle \langle \mathcal{T}o_3 o_4\rangle \nonumber \\
    &+ \langle \mathcal{T} o_1 o_3\rangle \langle \mathcal{T}o_2 o_4\rangle - \langle \mathcal{T} o_1 o_4\rangle \langle \mathcal{T}o_2 o_3\rangle \, .
\end{align}
This relation can be written in matrix form
\begin{align}
    \underline{\mathcal{G}}^{(2)}_c(1234) = \underline{\mathcal{G}}^{(2)}(1234) - {\underline{\chi}}^{=}_{0}(1234) - {\underline{\chi}}^{\times}_{0}(1234) \, ,
\end{align}
where ${\underline{\chi}}^{=}_{0}(1234)$ includes terms of the form $\langle \mathcal{T} o_1 o_2\rangle \langle \mathcal{T}o_3 o_4\rangle$, and ${\underline{\chi}}^{\times}_{0}(1234)$ includes terms of the form $\langle \mathcal{T} o_1 o_3\rangle \langle \mathcal{T}o_2 o_4\rangle$ and $\langle \mathcal{T} o_1 o_4\rangle \langle \mathcal{T}o_2 o_3\rangle$.
The full vertex $F$ can be computed from the connected part of the two-particle Green's function \cite{Negele1998, Rohringer2012}
\begin{align}
    G^{(2)}_{c}(1234) = - G(15)G(26) F(5678) G(73) G(84) \, . \label{eq:2P_decompose}
\end{align}

Due to momentum and energy conservation, and following the particle-hole convention \cite{Rohringer2012} of the Fourier transform, the momentum and frequency indices in the matrix above can be assigned as $k_1 = k$, $k_2 = k+q$, $k_3 = k'+q$, and $k_4 = k'$, with short hand notations $k=(K, i\omega_n)$ for fermionic and $q = (Q, i\nu_n)$ for bosonic indices.
Three independent spin combinations are possible in a SU(2) symmetric system, $\sigma_1 = \sigma_2 = \sigma_3 = \sigma_4$, $(\sigma_1 = \sigma_2) \neq (\sigma_3 = \sigma_4)$ and $(\sigma_1 = \sigma_4) \neq (\sigma_2 = \sigma_3)$, with $\sigma_j = \uparrow \text{or} \downarrow$ \cite{Rohringer2012}. 
We can then define
\begin{subequations}
\begin{align}
    \underline{\mathcal{G}}^{(2)}_{\sigma \sigma'}(kk'q) = \underline{\mathcal{G}}^{(2)}_{\sigma \sigma \sigma' \sigma'}(k,k+q,k'+q,k') \, , \label{eq:G4_uuud}
    \\
    \underline{\mathcal{G}}^{(2)}_{\overline{\sigma \sigma'}}(kk'q) = \underline{\mathcal{G}}^{(2)}_{\sigma \sigma' \sigma' \sigma}(k,k+q,k'+q,k') \, , \label{eq:G4_udbar}
\end{align}
\end{subequations}
where quantities in Eq.~(\ref{eq:G4_udbar}) can be obtained from those in Eq.~(\ref{eq:G4_uuud}) via SU(2) and crossing symmetries \cite{Rohringer2012}. 
Introducing  the full vertex matrix $\underline{\mathcal{F}}$, 
Eq.~\ref{eq:2P_decompose} can be written in matrix form as 
\begin{align}
    &\underline{\mathcal{G}}^{(2)}_{c, \sigma\sigma'}(kk'q) \\
    &= 
    - \frac{1}{\beta^2 N_c^2}\sum_{k_1 k_2}\underline{\chi}^{\times}_{0,\sigma\sigma}(kk_1q) ~ \underline{\mathcal{F}}_{\sigma \sigma'}(k_1k_2q) ~ \underline{\chi}^{\times}_{0,\sigma'\sigma'}(k_2k'q) \, . \nonumber
\end{align}
%


%

In order to extend fluctuation diagnostics to the superconducting state, we identify all independent scattering channels of the symmetry broken state and derive the corresponding equivalent expressions of the SD equation. 
In the basis of BdG spinors,
the creation of pairs of particles and holes in the spin singlet state is then described by the $4\times 4$ matrices $\mathit{\Sigma} = (i\sigma^y) \otimes \sigma^+$, $\bar{\mathit{\Sigma}} = (i\sigma^y) \otimes \sigma^-$ [$\sigma^{i = x,y,z}$ being the Pauli matrices, $\sigma^\pm = \frac{1}{2}(\sigma^x\pm i \sigma^y)$], while the corresponding terms for the triplet state are given by $T = \sigma^x\otimes \sigma^+$ and $\bar{T} = \sigma^x\otimes \sigma^-$.
The two $4\times4$ matrices $\rho =\mathbb{I}_{2\times 2}\otimes$ P$_p$ and $\bar{\rho} =\mathbb{I}_{2\times 2}\otimes $ P$_h$ [with P$_{p(h)} = \sigma^+\sigma^{-} (\sigma^-\sigma^+) $ the projector in the particle (hole) subspace] define the density operator of particles and holes. 
Analogously,  $S = \sigma^z\otimes$ P$_p$ and $\bar{S} = \sigma^z\otimes$ P$_h$ yield the spin operator, from which magnetic fluctuations originate.
With these definitions, the two-particle Green's function in different physical channels can be defined as
\begin{align}
    G^{O_a\,O_b}_{KK'Q}(&\tau_1, \tau_2, \tau_3, \tau_4) \nonumber \\
    &=
    \langle \mathcal{T} \hat{O}^{(a)}_{K, K+Q}(\tau_1, \tau_2) \hat{O}^{(b)}_{K'+Q, K'}(\tau_3, \tau_4) \rangle \, , \label{eq:two_part}
\end{align}
where the two time-dependent operators on the r.h.s.~are defined as  $\hat{O}^{(a)}_{K,K+Q}(\tau_1,\tau_2) = \Phi_{K}^\dagger(\tau_1) \cdot O_a \cdot \Phi^{\,}_{K+Q}(\tau_2)$, with $O_a$ corresponding to one of the eight 4$\times$4 matrices $\mathit{\Sigma}, \, \bar{\mathit{\Sigma}}; \, T, \, \bar{T}; \, \rho, \, \bar{\rho}; \, S, \, \bar{S}$.
All terms defined by Eq.~\ref{eq:two_part} can be computed from the linear combination of terms in $\underline{\mathcal{G}}^{(2)}_{c, \uparrow\uparrow}(kk'q)$ and $\underline{\mathcal{G}}^{(2)}_{c, \uparrow\downarrow}(kk'q)$.
The physical channels of the full vertex function can be defined with the same linear combinations as the two-particle Green's function.

The underlying symmetries of the system block-diagonalize the $8\times8$ matrix $G^{O_a\,O_b}$, such that each block identifies one independent scattering channel \cite{delre2021}.
In the paramagnetic state, where both the (global) $U(1)$ symmetry and the $SU(2)$ symmetry hold, only terms that conserve spin and particle number will be non-zero \cite{Rohringer2012}  giving rise to four independent scattering channels:
$\{\rho, \bar{\rho}\}$ define the density, $\{S, \bar{S}\}$ the magnetic, $\{\mathit{\Sigma}, \bar{\mathit{\Sigma}}\}$  the singlet-pairing, and $\{T, \bar{T}\}$ the triplet-pairing channel.

In the superconducting state, the spontaneously broken $U(1)$ symmetry allows for processes violating particle number conservation, and mix different scattering channels that are independent in the paramagnetic state.
However, the $SU(2)$ symmetry still holds and allows to identify two independent scattering channels: the density/spin-singlet channel $\{\rho,\bar{\rho},\mathit{\Sigma},\bar{\mathit{\Sigma}}\}$ and the magnetic/spin-triplet channel $\{S,\bar{S},T,\bar{T}\}$. 
Hence, in the superconducting phase Eq.~\ref{eq:SD} can be rewritten in two equivalent ways as 
\begin{align}
    \Sigma^{A}_{k} = \Sigma^{A, S}_{k} = \Sigma^{A, \rho}_{k} \, ,
    \label{eq:channel_sum}
\end{align}
with
\begin{widetext}
\begin{align}
    \Sigma^{A, S}_{k} &=
    \frac{1}{2}\frac{U}{(\beta N_c)^2} \sum_{k'q} \left[ G^{N}_{k+q} F_{\bar{T} S}(k k' q) G^{N}_{k'+q} G^{N}_{k'} +  G^{N}_{k+q} F_{\bar{T} T}(k k' q) G^{N}_{k'+q} G^A_{k'} \right] \nonumber \\
    &+\frac{1}{2}\frac{U}{(\beta N_c)^2} \sum_{k'q} \left[ G^{N}_{k+q} F_{\bar{T} \bar{T}}(k k' q) G^{A}_{k'+q} G^{N}_{k'} - G^{N}_{k+q} F_{\bar{T} \bar{S}}(k k' q) G^{A}_{k'+q} G^{A}_{k'} \right] \nonumber \\
    &-\frac{1}{2}\frac{U}{(\beta N_c)^2} \sum_{k'q} \left[ G^{A}_{k+q} F_{\bar{S} S}(k k' q) G^{N}_{k'+q} G^{N}_{k'} + G^{A}_{k+q} F_{\bar{S} T}(k k' q) G^{N}_{k'+q} G^{A}_{k'} \right] \nonumber \\
    &-\frac{1}{2}\frac{U}{(\beta N_c)^2} \sum_{k'q} \left[ G^{A}_{k+q} F_{\bar{S} \bar{T}}(k k' q) G^{A}_{k'+q} G^{N}_{k'} 
    - G^{A}_{k+q} F_{\bar{S} \bar{S}}(k k' q) G^{A}_{k'+q} G^{A}_{k'} \right]
    \, ,
    \label{eq:SD_Sz}
\end{align}
\begin{align}
    \Sigma^{A, \rho}_{k} &=
    \frac{1}{2}\frac{U}{(\beta N_c)^2} \sum_{k'q} \left[ G^{N}_{k+q} F_{\bar{\mathit{\Sigma}} \rho}(k, k', q) G^{N}_{k'+q} G^{N}_{k'} + G^{N}_{k+q} F_{\bar{\mathit{\Sigma}} \mathit{\Sigma}}(k, k', q) G^{N}_{k'+q} G^{A}_{k'} \right] \nonumber \\
    &-\frac{1}{2}\frac{U}{(\beta N_c)^2} \sum_{k'q} \left[ G^{N}_{k+q} F_{\bar{\mathit{\Sigma}} \bar{\mathit{\Sigma}}}(k, k', q) G^{A}_{k'+q} G^{N}_{k'} - G^{N}_{k+q} F_{\bar{\mathit{\Sigma}}\bar{\rho}}(k, k', q) G^{A}_{k'+q} G^{A}_{k'} \right] \nonumber \\
    &-\frac{1}{2}\frac{U}{(\beta N_c)^2} \sum_{k'q} \left[ G^{A}_{k+q} F_{\bar{\rho} \rho}(k, k', q) G^{N}_{k'+q} G^{N}_{k'} + G^{A}_{k+q} F_{\bar{\rho} \mathit{\Sigma}}(k, k', q) G^{N}_{k'+q} G^{A}_{k'} \right] \nonumber \\
    &+\frac{1}{2}\frac{U}{(\beta N_c)^2} \sum_{k'q} \left[ G^{A}_{k+q} F_{\bar{\rho} \bar{\mathit{\Sigma}}}(k, k', q) G^{A}_{k'+q} G^{N}_{k'} - G^{A}_{k+q} F_{\bar{\rho} \bar{\rho}}(k, k', q) G^{A}_{k'+q} G^{A}_{k'} \right] \, .
    \label{eq:SD_rho}
\end{align}
\end{widetext}
Eqs.~\ref{eq:SD_Sz} and \ref{eq:SD_rho} are the decompositions that enable application of the fluctuation diagnostics scheme to the superconducting state.
\section{Results}
\begin{figure}[tbh]
    \centering
    \includegraphics[width=0.9\columnwidth]{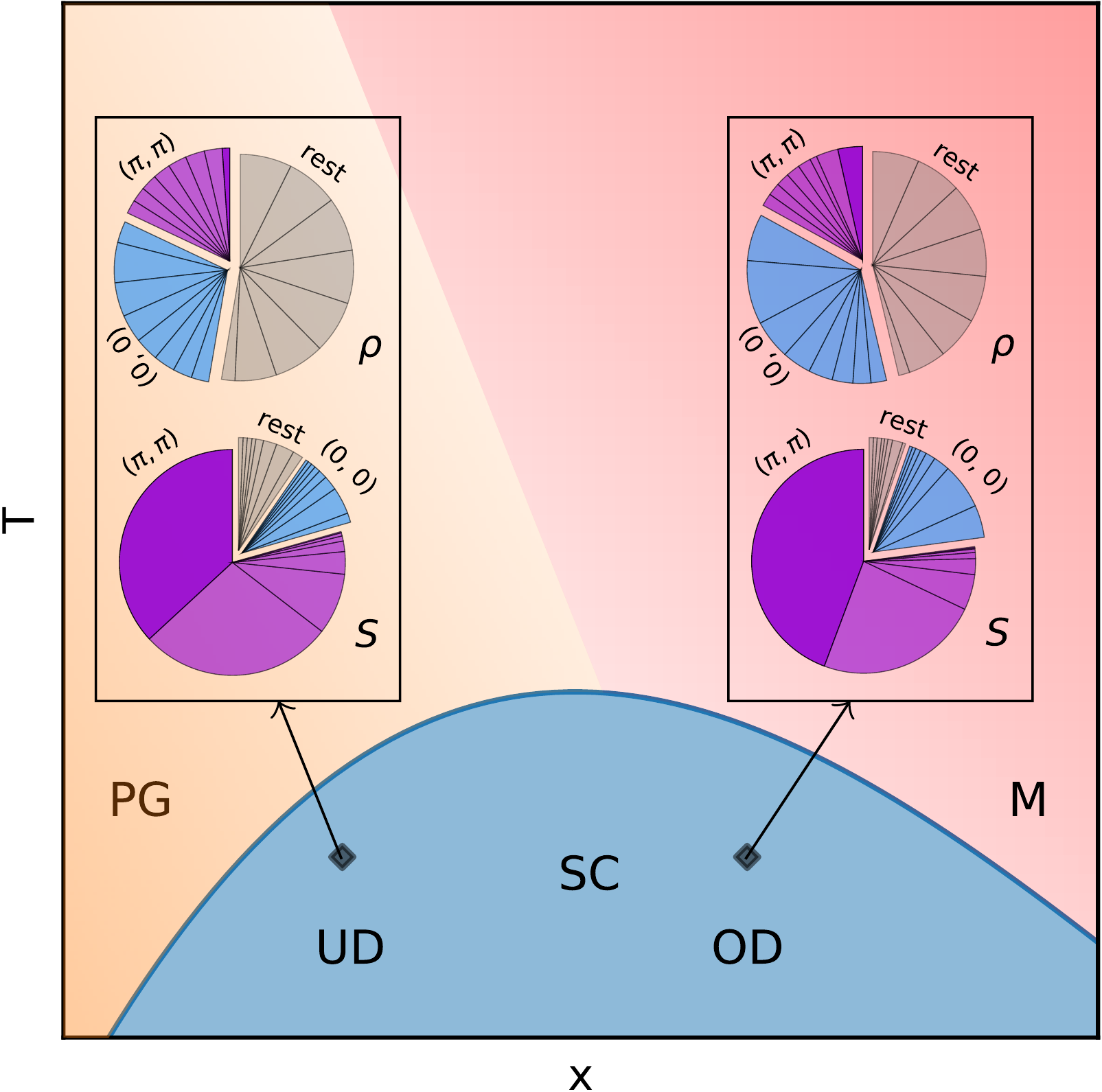}
    \caption{Phase diagram sketch, with pseudo-gap (PG, orange), metal (M, red), and superconducting (SC, blue) regimes. Black diamonds denote UD and OD data points analyzed in detail. Inset: Pie chart of $|\text{Re} \Sigma^{A}_{(\pi, 0), Q\nu} (i\omega_0)|$ in the density ($\rho$) and magnetic ($S$) channels. Counter-clockwise from the top, pieces represent contributions for momentum $Q = (\pi, \pi),~Q = (0, 0)$ and summation over the remaining momenta in an eight-site cluster. In each slice, separation indicates bosonic frequency $\nu_n$ with $n = 0, \pm1, ... \pm7$. }
    \label{fig:phase_diagram}
\end{figure}
\begin{figure}[tbh]
    \centering
    \includegraphics[width=0.9\columnwidth]{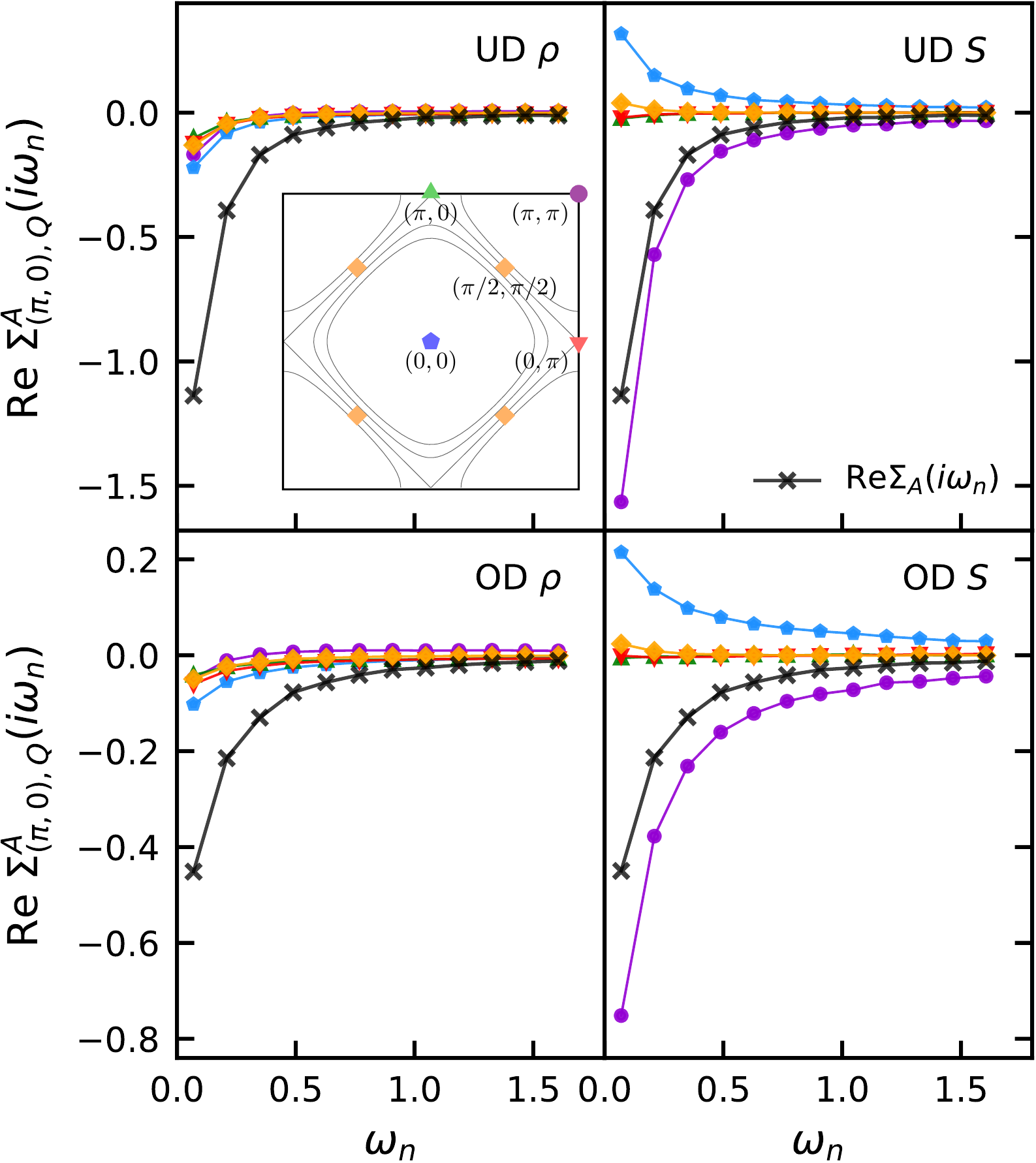}
    \caption{$\text{Re} \Sigma^{A}_{(\pi, 0), Q} (i\omega_n)$ for several transfer momenta $Q$ in the density ($\rho$) and magnetic ($S$) channels. OD and UD correspond to the diamonds in Fig.~\ref{fig:phase_diagram}. Black line with crosses: total anomalous self-energy after summation over all $Q$. Inset: non-interacting Fermi surface and location of momentum points corresponding to colors in main panels.
    }
    \label{fig:momentum}
\end{figure}

Fig.~\ref{fig:phase_diagram} gives the phase diagram of the 2D Hubbard model on the hole-doped side within the DCA approximation at intermediate interaction strength \cite{Gull2013} showing the pseudogap (PG), superconducting (SC), and metallic (M) regime. The pseudogap regime is characterized by a suppression of the single particle spectral function, and the superconducting phase corresponds to the region where the anomalous Green's function is non-zero. We present the results for two representative parameter sets without next-nearest neighbor hopping on an eight-site cluster with $U = 6t$, $\beta = 45t^{-1}$, i.e.~$x = 0.031$ ($T_c \in (t/30, t/35]$, corresponding to UD for this value of $U$) and $x = 0.075$ ($T_c \in (t/30, t/35]$, corresponding to OD for this value of $U$), see Ref.~\onlinecite{Gull2013} for a phase diagram. In DCA, both cases considered lie deep in the superconducting phase where
the anomalous Green's function $G^A_k$ is non-zero for $K = (0, \pi)$ and $(\pi, 0)$, with relation $G^A_{(0, \pi)} = -G^A_{(\pi, 0)}$. Cluster momentum points are shown in the inset of Fig.~\ref{fig:momentum}.
The inset of Fig.~\ref{fig:phase_diagram} shows the momentum ($Q$) and frequency ($\nu_n$) distribution of $|\text{Re} \Sigma^A_{(\pi, 0), Q\nu} (i\omega_0)|$, which is computed by summing over fermionic indices $k'$ but \textit{not} over $q$ in Eq.~\ref{eq:SD_Sz}, \ref{eq:SD_rho}. The pie chart insets show that for both UD and OD there is a dominant contribution from $Q = (\pi, \pi)$ and $\nu_n = 0$ in the magnetic/triplet channel $S$. In the density/singlet channel $\rho$, contributions from different momenta and frequencies are evidently distributed much more evenly.
\\
We first focus on the momentum distribution of $\text{Re} \Sigma^A_{(\pi, 0), Q} (i\omega_n)$ within the two physical channels in Fig.~\ref{fig:momentum}, computed by summing over all indices in Eq.~\ref{eq:SD_Sz}, \ref{eq:SD_rho} except for the transferred momentum $Q$.
The inset in the upper left panel shows the momentum points in an eight-site cluster, and the Fermi surface in the non-interacting system for dopings of 0.2 (corresponding to hole doping), 0, -0.2, and -0.4 (corresponding to electron doping).
The left two panels show the contribution of different $Q$ in the density channel. The weak $Q$ dependence indicates the absence of a dominant mode in this channel.
Results for the magnetic/triplet channel are shown in the right two panels. The transfer momentum $Q = (\pi, \pi)$ associated with AFM fluctuations is clearly the \textit{dominant} mode in both the UD and the OD regime. We note that a sub-leading, though still sizable,  {\sl negative} contribution to the anomalous self energy is originated by a ferromagnetic mode with $Q = (0, 0)$.
The black lines with crosses are computed by summing over all different momenta in the cluster, resulting in $\Sigma^{A, S}$ and $\Sigma^{A, \rho}$ of Eq.~\ref{eq:channel_sum}.
\\
\begin{figure}[tbh]
    \centering
    \includegraphics[width=0.9\columnwidth]{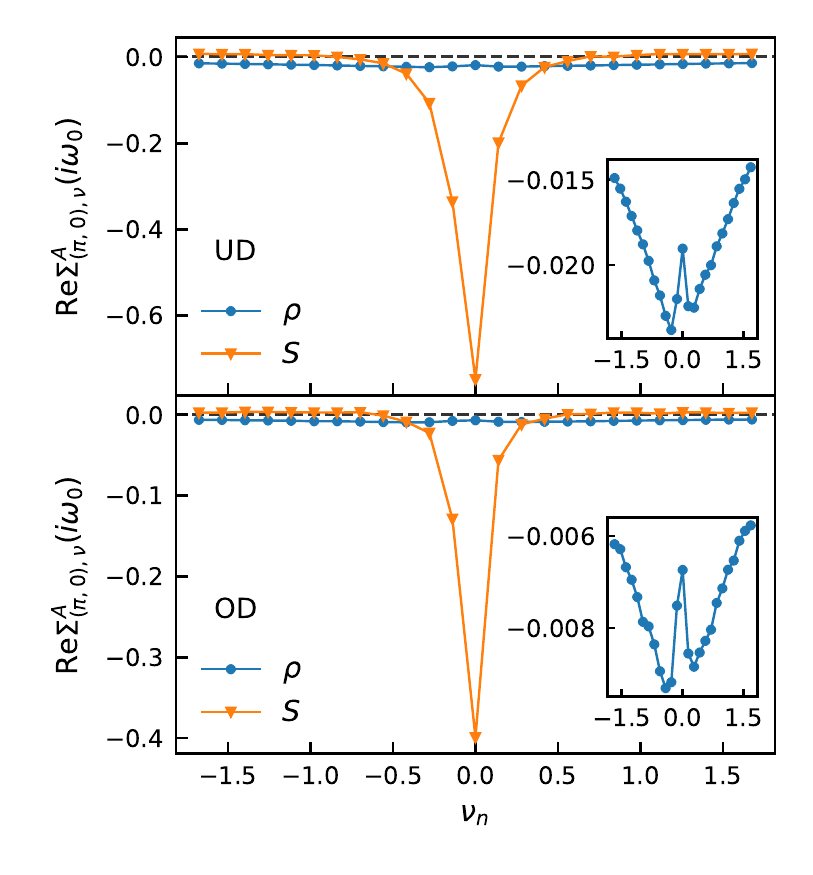}
    \caption{$\text{Re} \Sigma^{A}_{(\pi, 0), \nu} (i\omega_0)$ in density ($\rho$) and magnetic ($S$) channels. OD and UD correspond to the two data points of Fig.~\ref{fig:phase_diagram}. Inset: $\text{Re} \Sigma_{A, \nu} (i\omega_0)$ in density channel with rescaled $y$ axis.}
    \label{fig:frequency}
\end{figure}
Important insight can be gained by a complementary analysis in frequency space: Fig.~\ref{fig:frequency} shows the frequency dependence of $\text{Re} \Sigma^A_{(\pi, 0), \nu_n} (i\omega_0)$, corresponding to the result at the lowest fermionic Matsubara frequency $i\omega_0 = {\pi}/{\beta}$ after summation over all indices except for bosonic frequency $\nu_n$ in Eq.~\ref{eq:SD_Sz}, \ref{eq:SD_rho}.
The low-frequency peak in the magnetic channel identifies the corresponding fluctuation as a well-defined and long lived ``mode''. In the density channel, the contributions are more evenly distributed across a wide frequency range, corresponding to short lived fluctuations.

\section{Discussion and Conclusion}
By extending the fluctuation diagnostics approach to the superconducting phase, we have been able to unambiguously identify spin-fluctuations \cite{Scalapino1986, Miyake1986, Scalapino1999, Scalapino2012} as the dominant contribution to the $d-$wave pairing in the Hubbard model at interaction strengths believed to be relevant for the cuprates, i.e.,~\textit{beyond} the weak-coupling regime \cite{Zanchi2000, Raghu2010,Raghu2010, Scalapino1986,Deng2015}.
At the same time, consistent with the existing work in the normal state \cite{Gunnarsson2015,Wu2017,Dong2019}, we do \textit{not} find any indication supporting the alternative scenarios mentioned in the introduction, such as nematic fluctuations \cite{Fradkin2010}, loop current order \cite{Allais2012}, or ``intertwining'' of different orders \cite{Fradkin2015}. In the latter case, multiple fluctuations such as density or magnetic ones would contribute synergistically to the pairing, rather than compete, in contradiction with our results.
We emphasize that the fluctuation diagnostic is capable of detecting the occurrence of this situation, when it is realized, e.g.~in the attractive Hubbard model \cite{Gunnarsson2015}. 

Our identification of the superconducting glue agrees with the findings of several experiments. Ref.~\onlinecite{Dahm2009} finds good quantitative agreement between the spectral function computed from conventional spin fluctuation theory with magnetic susceptibility measured by inelastic neutron scattering, and the spectral function measured from angle-resolved photoemission spectroscopy in the superconducting phase of YBCO. 
Inelastic photon scattering experiments \cite{Wang2020} on Hg1201 and Hg1212 infer that the superconducting temperature $T_c$ can be determined by the strength of the magnetic interactions (``paramagnon signals"), supporting the theory of magnetically mediated high-temperature superconductivity.
Other experiments suggest a relation between superconductivity and charge density wave \cite{chu2021}, or that the pseudogap and superconductivity may have different origins \cite{Wu2020}. Thus, the numerical findings of our study suggest the possibility that the latter class of experiments may be probing aspects of cuprates physics beyond those encoded in the single-orbital Hubbard model on an eight-site DCA cluster.

Independently of the agreement with this multifaceted experimental evidence, our identification of the superconducting glue in terms of spin-fluctuations touches a delicate and  important aspect of the theoretical description of high-$T$ superconductivity.
In particular, conventional spin fluctuation theory as described in Refs.~\onlinecite{Maier2006, Maier2007, Maier2008, Maier2019} appears only able to capture a fraction of the pairing contribution \cite{Dong2022}. 
The origin of this discrepancy can be ascribed to the RPA-like one-loop spin fluctuations expressions used in conventional approaches which, outside of the weak-coupling regime, do not capture all spin-fluctuation mediated processes \cite{Kitatani2019}. 

The microscopic picture of superconductivity emerging from our analysis agrees well with recent studies of the description of the \textit{non}-superconducting pseudogap regime: While spin-fluctuations were identified as the predominant mechanism of the pseudogap \cite{Gunnarsson2015, Wu2017, Dong2019}, differences with respect to the predictions of conventional spin-fluctuation theory were found and traced \cite{Krien2021}  to the imaginary part of the dynamical scattering amplitude between electrons and spin fluctuations, which is absent in conventional approaches \cite{Krien2021}.

In conclusion, our fluctuation diagnostics of the superconducting order in the Hubbard model precisely identifies antiferromagnetic spin fluctuations as the glue of the $d-$wave pairing. This conclusion applies to the intermediate-to-large values of the electronic interaction relevant to cuprate physics. For this reason, the  spin-fluctuations-driven pairing found in our calculations are expected to differ from conventional spin-fluctuation theories.  %

\acknowledgments{
XD and EG are supported by NSF DMR 2001465. LD  acknowledges DE-SC0019469.
AT acknowledges financial support from the Austria Science Fund (FWF) through the Project I 5868-N – FOR 5249 (QUAST).
We thank A.J. Millis,  Kai Sun and F.~Krien for detailed and insightful discussions. 
}

\bibliography{refs.bib} 

\end{document}